\documentclass[12pt]{article}

\usepackage{amsmath,amssymb,amsbsy,amsfonts,latexsym,graphicx}
\usepackage{hyperref}
\usepackage{color, array, subfigure}
\usepackage{cite}
\usepackage{lipsum}

\allowdisplaybreaks

\evensidemargin \oddsidemargin

\newcommand\blfootnote[1]{%
  \begingroup
  \renewcommand\thefootnote{}\footnote{#1}%
  \addtocounter{footnote}{-1}%
  \endgroup
}

\begin{document}

\begin{titlepage}



\vspace{15mm}
\baselineskip 9mm
\begin{center}
  {\Large \bf 
{  
Construction of Superconducting Dome and Emergence of Quantum Critical Region in Holography} 
 }
\end{center}

\baselineskip 6mm
\vspace{10mm}
\begin{center}
Yunseok Seo$^{a}$, Sejin Kim$^{b}$,  and Kyung Kiu Kim$^{c}$
 \\[10mm] 
  {\sl College of General Education, Kookmin University, Seoul 02707, Korea}
   \\[3mm]
\blfootnote{{\tt  ${}^a$yseo@kookmin.ac.kr, ${}^b$sejin817@kookmin.ac.kr, ${}^c$kimkyungkiu@kookmin.ac.kr}}
  
\end{center}

\thispagestyle{empty}

\vspace{1cm}
\begin{center}
{\bf Abstract}
\end{center}
\noindent
In this work, we investigate an extended model of holographic superconductor by a non-linear electrodynamic interaction coupled to a complex scalar field. This non-linear interaction term can make a quantum phase transition at zero temperature with finite charge carrier density. By solving full equations of motion, we can construct various shapes of the superconducting phase in the phase diagram. With a specific choice of interaction coefficients, we can construct a phase diagram with a superconducting dome.  Also, we find a new geometric solution inside the superconducting dome, which turns out to be a Lifshitz-type geometry. This geometry is characterized by a dynamical critical exponent, which plays a crucial role near the quantum critical point. We refer to this region in the phase diagram as a `quantum critical region.'
\\ [15mm]

\vspace{5mm}
\end{titlepage}

\tableofcontents
\section{Introduction}
A high $T_C$  superconductor is one of the most fascinating materials due to its various applications in the industry and theoretical perspective. For the conventional superconductor, BCS\cite{bardeen1957microscopic} theory can explain superconducting phase transition by using a pairing mechanism. However, this theory is imperfect in understanding the high $T_C$ superconducting phase transition. Therefore, constructing the theoretical framework of the high $T_C$ superconductor is still challenging in condensed matter physics.

The many high $T_C$ superconducting materials have various phases. For example, the high $T_C$ cuprate shows several phases in doping parameters and temperature \cite{vishik2018photoemission}. Together with the superconducting dome, the phase diagram contains an anti-ferromagnetic insulating phase, a strange metal phase, and a Landau-Fermi liquid phase in the phase diagram. These various phases are believed to be raised by the strongly correlated nature of the electron system. Therefore, we need a theoretical tool beyond perturbative calculations to understand these phases and their transition phenomena.

Similar to other materials with several phases, there is a quantum critical point at zero temperature. In the high $T_C$ cuprate case, this quantum critical point is hidden behind the superconducting dome phase. Due to the absence of thermal fluctuation, the transition at the quantum critical point is believed to be fully quantum. 

A condensed matter study says that there are new scaling symmetries near the quantum critical point, and these symmetries are classified by two critical exponents \cite{senthil2004deconfined,sachdev2010quantum,shibauchi2014quantum,polkovnikov2005universal,schlief2017exact,abanov2004anomalous}. One exponent is called a `dynamical critical exponent' $z$. The dynamical critical exponent indicates the different scaling of time and space, $ x \rightarrow \lambda x$, $t \rightarrow \lambda^z t$. The other exponent is the `hyperscaling violation exponent' $\theta$, which means that the effective theory can be described in $d-\theta$ spatial dimensions.

Due to the strongly interacting nature of the high $T_C$ superconducting materials, it is extremely hard to understand the phase transition mechanism and properties near the quantum critical point by traditional approaches. Fortunately, gauge/gravity duality based on AdS/CFT correspondence \cite{Maldacena:1997re,Gubser:1998bc,Witten:1998qj} has been widely studied for understanding strongly interacting systems \cite{Hartnoll:2007ih,Seo:2016vks,Kim:2014bza,Kim:2015wba,Seo:2015pug,Seo:2017oyh,Seo:2018hrc,Kim:2019lxb,Kim:2020ozm,Seo:2023bdy}. A gravity model describing superconductors was first introduced by  Hartnoll, Herzog and Horowitz(HHH) \cite{Hartnoll:2008vx,Hartnoll:2008kx}. After HHH's work, enormous studies and developments have been achieved by many researchers. We refer \cite{Hartnoll:2009sz,Herzog:2009xv,Horowitz:2010gk} for reviews and references.

The original model of the holographic superconductor has a translation symmetry in the theory, and hence, there is a delta function peak at $\omega=0$ in the electric conductivity. This delta function exists even in the normal phase.  Therefore, whether this delta function peak comes from a superconducting nature or the translation symmetry is unclear. This issue can be resolved by introducing a linear axion field proposed for momentum relaxation in the boundary theory \cite{Andrade:2013gsa,Baggioli:2021xuv}. In the work of \cite{Kim:2015dna,Kim:2016hzi}, they confirmed that the delta function peak in the electric conductivity appears only in the superconducting phase, while the Drude like peak is in the normal phase.

The other issue in the holographic superconductor is that it is hard to construct a dome-shaped superconducting phase. In the original model of the holographic superconductor, the existence of scalar hair can be understood from the Breitenlohner-Freedman (BF) bound of the model. The BF bound is a bound of the scalar mass such that the corresponding scaling dimension to be real. The BF bound in the $AdS_{d+1}$ spacetime is determined as $m^2_{BF} \ge - d^2/4$. In the zero temperature(extremal) limit, the near horizon geometry of 4 dimensional RN black hole becomes $AdS_2 \times S^2$ with a different $AdS$ scale, while the boundary geometry is $AdS_4$. Therefore, BF bound changes along the radial direction at zero temperature. In a certain range of scalar mass, the BF bound can be violated near the horizon while it satisfies the bound at the boundary. It makes instability near the horizon; hence, the black hole geometry with scalar hair can be formed. In the original model of the holographic superconductor \cite{Hartnoll:2008vx}, however, the BF bound analysis only depends on the mass of the scalar field. Therefore, once the scalar mass is determined such that scalar hair appears, the superconducting phase covers the whole range of the charge carrier density at the zero temperature.

To resolve this problem, authors in \cite{Kiritsis:2015hoa} introduce a non-Abelian gauge field, triplet scalar field, and two $U(1)$ gauge fields. By tuning the order parameters corresponding to each field, they study the competition between order parameters and get a rich phase diagram structure. There is a similar problem of infinite DC conductivity in the metallic phase due to the translation invariance. This issue is resolved in \cite{Baggioli:2015dwa} by introducing a linear scalar field, which leads to momentum relaxations. However, the existence of the quantum criticality in the superconducting phase is not clear in their study. The authors in \cite{Kim:2015dna}, introduce a linear axion field to relax the momentum conservation. They found that there is a delta function peak in the electric conductivity at zero frequency in the superconducting phase in the presence of the momentum relaxation. They also observed the possibility of the quantum phase transition in a certain range of scaling dimensions of the complex scalar field.
 
 Recently, there have been works on the holographic superconductor model in the context of fermion's Green's function \cite{Ghorai:2023wpu,Ghorai:2023vuo}, in the Gubser-Rocha model \cite{Zhao:2023qms} and in the $f(R)$ gravity \cite{Roussev:2023uyo}. 
 
 In this paper, we construct a model of a holographic superconductor with non-linear electrodynamic interactions between a charged scalar field and $U(1)$ gauge field. In the previous work \cite{Seo:2023bdy}, we find that the non-trivial interaction between the scalar field and the bulk $U(1)$ gauge field can lead to a quantum phase transition at the finite charge density and the zero temperature. Inspired by \cite{Seo:2023bdy,Baggioli:2016oju}, we extend this model to the complex scalar model, which describes the holographic superconducting phase. We also introduce non-linear electrodynamic interaction to the complex scalar field for making a superconducting dome in the phase diagram.

We solve coupled equations of motion with proper boundary conditions and find hairy black hole solutions in a wide range of non-linear electrodynamic interaction coefficients. We also get various shapes of the superconducting phase in the phase diagram. At a certain choice of parameters, we can construct a superconducting dome phase.

We also investigate properties inside the superconducting phase. As expected, the value of condensation, which is proportional to the superconducting order, increases as temperature increases. However, we find certain regions inside the superconducting dome where hairy black hole solution does not exist. We analyze equations of motion near the boundary and find that, interestingly, new geometry appears in this region. The new geometry turns out to be a Lifshitz type geometry. This geometry is parametrized by a dynamical critical exponent that indicates different scaling of time and space. The dynamical critical exponent $z$ plays a crucial role near the quantum critical point. Therefore, we speculate this region corresponds to the `quantum critical region'.

The organization of the paper is as follows: In section 2, we construct a model of a holographic superconductor with non-linear electrodynamic interactions with a charge scalar field. We analyze the thermodynamic variables of the model. In section 3, we study the possibility of a quantum phase transition and construction of the superconducting dome using a BF bound analysis. In section 4, we get various superconducting phase diagrams by solving equations of motion numerically with proper boundary conditions. We also find a new geometric solution with Lifshitz scaling symmetry inside the superconducting phase. In section 5, we summarize our work and discuss the future directions.
 
\section{Holographic Model coupled to Non-linear Electrodynamics}


In this section, we propose a model which is appropriate to describe a nontrivial structure of the superconducting phase. We begin with the Einstein-Maxwell-Axion model with a complex scalar field to describe the superconducting phase with momentum relaxation.   The condensation of the scalar field without source leads to the superconducting phase, which is the usual holographic superconductor model. As discussed in the previous section, we introduce non-linear electrodynamic interaction between the $U(1)$ gauge field and the complex scalar field.  The starting action of the model is as follows;

\begin{align}\label{S0}
S_{B} = \frac{1}{16\pi G} \int d^4 x\sqrt{-g}\left( R+ \frac{6}{L^2} - \frac{1}{2}\sum_{i=1}^2\left(\partial\chi^{i}\right) - \frac{1}{4}F^2 -|D\varphi|^2 - m^2 |\varphi|^2\right) + S_{NED}\,, 
\end{align}
where the non-linear electrodynamic action is set as
\begin{align}\label{SNED}
S_{NED} = - \frac{1}{16\pi G} \int d^4 x \sqrt{-g}\, \frac{1}{4}  |\varphi|^2\left(\gamma_2 F^2 +\gamma_4 L^2 (F^2)^2\right)\,.
\end{align}
Here, we introduce the linear axion field $\chi_i$ and the complex scalar field $\varphi$ to realize an impurity effect and a superconducting order, respectively. Also, the electromagnetic field is given $F=dA$ and $F^2$ denotes $F_{MN}F^{MN}$. 

From the action (\ref{S0}), we can get equations of motion. The Einstein equations are given by
\begin{align}\label{EinsteinEQ}
&R_{M N} -\frac{1}{2}g_{M N} {\cal L} -\frac{1}{2}\sum_{i=1}^{2} \partial_M \chi^i \partial_N \chi^i -\frac{1}{2}\left(1+\gamma_2 |\varphi|^2\right)F_M{}^P F_N{}_P \cr
&- \frac{1}{2}\left(D_M\varphi (D_N\varphi)^*+D_N\varphi (D_M\varphi)^*\right)-\gamma_4 L^2 \,|\varphi|^2 F_M{}^P F_{NP} F^2=0\,,
\end{align}
where ${\cal L}$ is a Lagrangian density of the action $S_B$ and the covariant derivative is given by $D_N \varphi = \left(\nabla_N - i e A_N\right)\varphi$, where $e$ is the  electric charge of the complex scalar field.

The Maxwell equations can be written as

\begin{align}\label{MaxwellEQ}
\nabla_M \left( 1 + \gamma_2 |\varphi|^2 + 2  \gamma_4 L^2\,|\varphi|^2 F^2\right) F^{MN}  = i e \left(\varphi^* D^N \varphi - \varphi (D^N \varphi)^* \right)\,.
\end{align}
Notice that no conserved charge can be defined due to the complex scalar field. Therefore, we have to get the charge density as a conjugate value of the chemical potential, which is the boundary value of the $U(1)$ gauge field $A_t(r)$. We will discuss it later.

The equation of motion of the axion field becomes
\begin{align}\label{AxionEQ}
\nabla^2\chi^i = 0.
\end{align}
Finally, the equations of motion for the complex scalar field are
\begin{align}\label{ScalarEQ}
&D^2 \varphi -\left(m^2+\frac{1}{4}\gamma_2 F^2 +\frac{1}{4} \gamma_4 L^2\,(F^2)^2 \right) \varphi = 0 \cr
&(D^2 \varphi)^* -\left(m^2+\frac{1}{4}\gamma_2 F^2 +\frac{1}{4} \gamma_4 L^2\,(F^2)^2 \right)\varphi^* = 0
\end{align}
The  We adopt the coordinate indices as $x^M=(x^\mu,r)$ and $x^\mu=(t,x^i)=(t,x,y)$. Choosing a gauge $\nabla_M A^M=A_r = 0$, the scalar field equation can be written as follows:
\begin{align}
\left(\nabla^2- e^2 A_\mu A^\mu - m^2 -\frac{1}{4}\left(\gamma_2 F^2 + \gamma_4 L^2\,(F^2)^2 \right) \right)\varphi (\varphi^*)= 0\, .
\end{align}
Here, one can read the effective mass as
\begin{align}\label{Meff}
m^2_{\text{eff}}= m^2+ e^2 A_\mu A^\mu  +\frac{1}{4}\left(\gamma_2 F^2 + \gamma_4 L^2\,(F^2)^2 \right)\,.
\end{align}

Now, we take a suitable ansatz for a hairy black brane solution as follows:
\begin{align}\label{Ansatz}
&ds^2 = - U(r) e^{2(W(r)-W(\infty))} dt^2 + \frac{r^2}{L^2} \left(dx^2 + dy^2\right) + \frac{dr^2}{U(r)}\nonumber\\
&\chi^i = \kappa (x,y)~,~A= A_t(r) dt~,~\varphi = \phi(r) e^{i \tilde{\phi}(r)}\,,
\end{align} 
where  $\phi$ and $\tilde{\phi}$ are real and imaginary part of $\varphi$.

By imposing the metric ansatz(\ref{Ansatz}) into the equations motion(\ref{EinsteinEQ})-(\ref{ScalarEQ}), we get equations of motion for each fields as follows:
\begin{align}\label{eom1}
\phi ''&+\frac{\phi ' \left(r W'+2\right)}{r}+\frac{U' \phi '}{U}-\frac{m^2 \phi }{U}+\frac{e^2 e^{-2 W} \phi  a_t^2}{U^2}+\frac{\gamma_2  e^{-2 W} \phi  \left(a_t'\right){}^2}{2 U}\nonumber\\
&-\frac{\gamma_4  L^2 e^{-4 W} \phi  \left(a_t'\right){}^4}{U}=0\\
a_t''&-\frac{2 e^2 e^{2 W} \phi ^2 a_t}{U\, {\cal F}}+\frac{4 \gamma_4  L^2 \phi ^2 \left(a_t'\right){}^3 \left(3 r W'-2\right)}{r {\cal F}}-\frac{e^{2 W} \left(\gamma_2  \phi ^2+1\right) a_t' \left(r W'-2\right)}{r {\cal F}}\cr
&+\frac{2 \phi  \phi ' a_t' \left(\gamma_2  e^{2 W}-4 \gamma_4  L^2 \left(a_t'\right){}^2\right)}{{\cal F}}=0\label{eom2}\\
U'&+\frac{1}{4} r e^{-4 W} \phi ^2 \left(-6 \gamma_4  L^2 \left(a_t'\right){}^4+\gamma_2  e^{2 W} \left(a_t'\right){}^2+2 m^2 e^{4 W}\right) \nonumber\\
&+\frac{e^2 r e^{-2 W} \phi ^2 a_t^2}{2 U}+\frac{1}{4} r e^{-2 W} \left(a_t'\right){}^2+\frac{\kappa ^2 L^2}{2 r}-\frac{3 r}{L^2}+\frac{U \left(r^2 \left(\phi '\right)^2+2\right)}{2 r}=0\label{eom3}\\
W'&-\frac{1}{2} r \left(\phi '\right)^2-\frac{e^2 r \phi ^2 e^{-2 W} a_t^2}{2 U^2}=0\,,\label{eom4}
\end{align}
where $a_t=e^{W(\infty)}A_t$ and ${\cal F}\equiv  \phi ^2 \left(\gamma_2  e^{2 W}-12 \gamma_4  L^2 \left(a_t'\right){}^2\right)+e^{2 W}$. Notice that the only real part of the scalar field appears in the equations of motion by fixing the phase of the field. The linear axion field automatically satisfies the equation of motion(\ref{AxionEQ}).

It is natural to take a scaling of fields and parameters for a numerical method. The proposed scaling is summarized as follows:
\begin{align}\label{DimLess}
&r=r_h\tilde{r},\,U(r) = \frac{r_h^2}{L^2}\tilde{U}(\tilde{r}),\,e=\frac{\tilde{e}}{L},\,a_t(r)=\frac{r_h}{L}\tilde{a}_t(\tilde{r}),\nonumber\\
&\kappa=\frac{r_h}{L^2}\tilde{\kappa},\,m^2=\frac{\tilde{m}^2}{L^2},\,W(r)=\tilde{W}(\tilde{r})\,,\varphi(r)=\varphi(\tilde{r})\,,
\end{align}
where the tilde denotes dimensionless quantities. An efficient way to deal with (\ref{eom1})-(\ref{eom4}) with this scaling is to take $L=1$ and regard all quantities and fields as dimensionless ones.
 
In the presence of the black hole horizon, the temperature and the entropy density of the boundary system are defined as
\begin{align}\label{TS}
T= \frac{1}{4\pi} U'(r_h) e^{W (r_h)-W (\infty)}= \frac{r_h}{4\pi L^2}\tilde{U}'(1)e^{\tilde{W}(1)-\tilde{W}(\infty)}~,~s =\frac{r_h^2}{4GL^2}\,.
\end{align}
From the equations of motion, we get the relation between on-shell action and other quantities as
\begin{align}
\Omega_{os}  = \frac{\mathcal{V}}{16\pi G} \left( (2 M-\mu Q) L^2 - \frac{r_h^2}{L^2} e^{W(r_h)-W(\Lambda)} U'(r_h) \right)\,,
\end{align}
where ${\mathcal V}$ is a volume of transverse directions. If we identify the on-shell action with the negative pressure $\Omega_{os}/{\cal V}\equiv - {\cal P}$, we get the Smarr relations;
\begin{align}
{\cal P} + \epsilon = \mu Q + s T,
\end{align}
here, we use (\ref{TS}) and set $16 \pi G = L = 1$. The energy density can be obtained from the energy-momentum tensor as
\begin{align}
\epsilon = \left<T_{00} \right> = 2 M.
\end{align}
The details of the calculations are shown in Appendix \ref{OnS}.

\section{BF bound Analysis}

In this section, we study which parameter space allows a phase transition between the RN-AdS black brane and the hairy black brane. The Breitenlohner-Freedman (BF) bound is usually used to find the possible parameter space. The BF bound is the lower bound of the mass of the scalar field not to be tachyonic. The BF bound usually depends on the spacetime dimension and form of the field. Then, the difference of the geometry at the boundary and the near horizon of the extremal black hole causes instability by violating the BF bound, and hence, a hairy black hole geometry can be possible.

We start with the extremal black brane, whose metric is
\begin{align}
U(r)=\frac{\left(r-r_h\right){}^2}{L^2} \left(1+\frac{2 r_h}{r}+\frac{3 r_h^2}{r^2}-\frac{\kappa ^2 L^4}{2 r^2}\right)\,,
\end{align}
where the horizon radius is determined as $r_h^2 =\frac{\kappa^2  L^2}{12} \left(1+\sqrt{1+12 \frac{Q^2}{\kappa ^4}}\right)$ from the extremal condition. The near horizon limit of the metric becomes 
\begin{align}
U(r)\sim \frac{(r-r_h)^2}{L^2} \left( \frac{6 \sqrt{1+ 12 Q^2/\kappa^4}}{1+\sqrt{1+ 12 Q^2/\kappa^4} } \right)\,.
\end{align}
This geometry is nothing but $AdS_2\times \mathbb{R}^2$ with the $AdS$ radius,
\begin{align}
L_{\text{eff}}^2= \frac{L^2}{6}\left( 1 + \frac{1}{\mathbf{p}} \right)\,,
\end{align}
where $\mathbf{p}=\sqrt{1+12 Q^2/\kappa^4}$.

The BF bound of a scalar field in $(d+1)$-dimensional $AdS$ spacetime with a radius $L$ is given by $m_{BF}^2=-d^2/4 L^2$, so the mass of the scalar field $m^2$ is always greater than $-9/4 L^2$ for the asymptotically stable $AdS_4$. Due to the coupling to the $U(1)$ gauge field, the mass of the scalar field changed to the form of the effective mass (\ref{Meff}). In the effective mass formula, the gauge potential $F_{\mu\nu}$ falls off at the $AdS$ boundary, and then the effective mass at the boundary is the same as the scalar field's mass.

  On the other hand, the gauge potential does not vanish at the horizon. The effective mass (\ref{Meff}) of the near horizon limit turns out to be 
\begin{align}\label{HRMeff}
\lim_{r\to r_h} m^2_{\text{eff}}=m_h^2=m^2-2 e^2 \left(1-\frac{1}{\mathbf{p}}\right)-\frac{6 \gamma_2 }{L^2}\frac{(\mathbf{p}-1)}{(\mathbf{p}+1)} + \frac{12^2 \gamma_4 }{L^2}\frac{(\mathbf{p}-1)^2}{(\mathbf{p}+1)^2}
\,.
\end{align}
The second term in (\ref{HRMeff}) is lowering the effective mass of the scalar field which appears in the usual holographic superconductor model. The third and the last term contribute oppositely to the effective mass. 
 In addition, a stability condition near the horizon is
\begin{align}
m_h^2 L_{\text{eff}}^2 > - 1/4\,,
\end{align}
which is explicitly
\begin{align}\label{BFcond}
\mathbb{H}=\frac{1}{6}\left(1+\frac{1}{\mathbf{p}}\right)\left\{ \tilde{m}^2-2 \tilde{e}^2 \left(1-\frac{1}{\mathbf{p}}\right)-\frac{6 \gamma_2 }{L^2}\frac{(\mathbf{p}-1)}{(\mathbf{p}+1)} + \frac{12^2 \gamma_4 }{L^2}\frac{(\mathbf{p}-1)^2}{(\mathbf{p}+1)^2}  \right\} > -\frac{1}{4}\,,
\end{align}
where the dimensionless mass and charge, $\tilde{m}$ and $\tilde{e}$, are given in (\ref{DimLess}). The violation of this condition could give rise to a hairy configuration.

\begin{figure}[ht!]
\begin{center}
\subfigure[$\gamma_2 =\gamma_4 =0$]
   { \includegraphics[width=5cm]{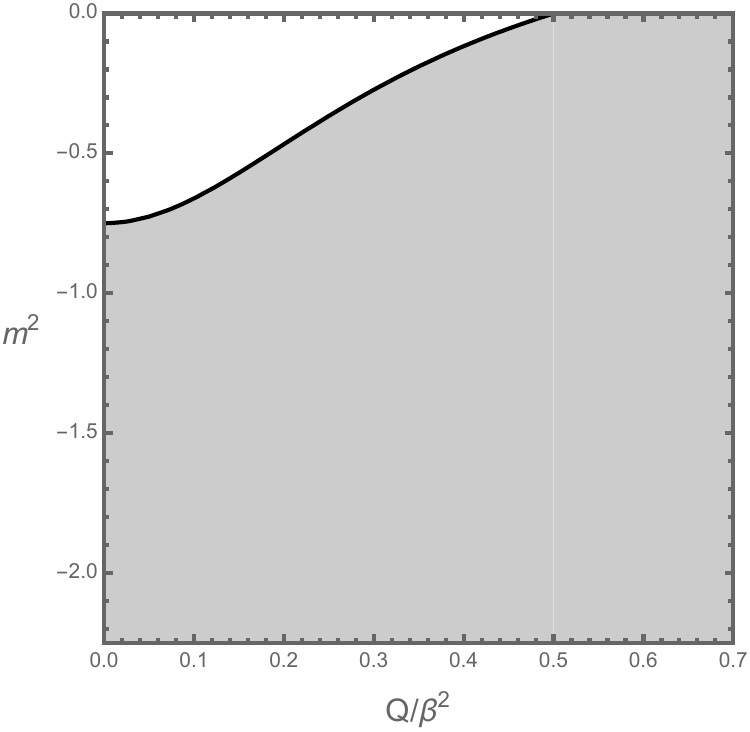}  }
   \hskip1.5cm
 \subfigure[$\gamma_2=6,~\gamma_4=1.5$]
   { \includegraphics[width=5cm]{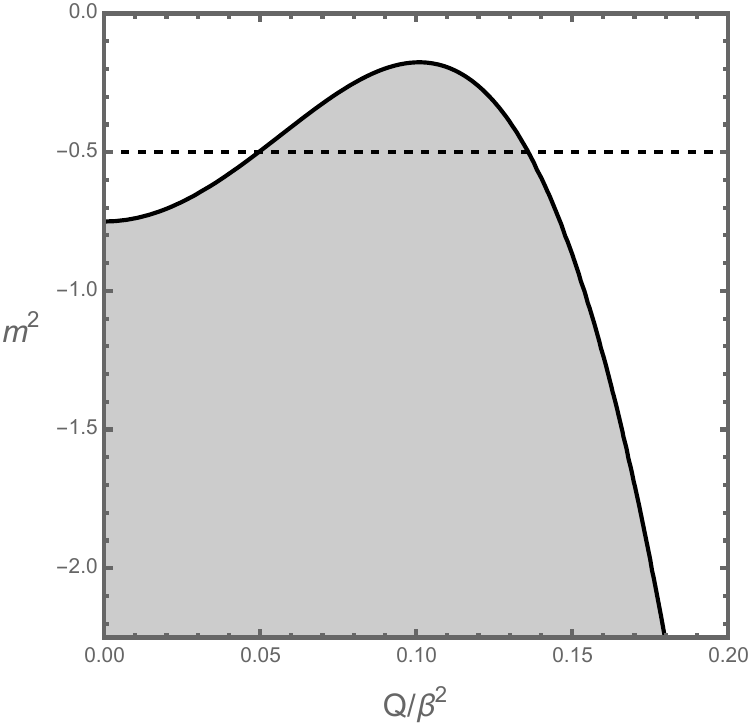}  }

\caption{BF bound analysis without(a) and with(b) non-linear electrodynamics interactions. We fix the scalar charge to $e =1$ in both case.}
\label{fig:BF}
\end{center}
\end{figure}

Some examples of  BF bound analysis(\ref{BFcond}) are shown in Figure \ref{fig:BF}.

Figure \ref{fig:BF} (a) shows BF bound analysis for the conventional holographic superconductor model \cite{Hartnoll:2008vx}. The shaded region in the figure denotes the region of the violating $AdS_2$ BF bound near zero temperature. In the absence of the charge density $Q$, the region where $m^2 > -3/4$ is does not allow the scalar field condensation even in the presence of the $\gamma_2$, $\gamma_4$ coupling. In this region, the unitarity of the scalar field is violated; hence, the scalar field cannot be condensed. As shown in the figure, near horizon BF bound is always violated when $m^2 =-2$, therefore corresponding superconducting phase possibly exist near zero temperature in all charge density range. It is consistent with numerical calculations \cite{Kim:2015dna}?.

This phenomenon changes drastically in the presence of non-linear electrodynamic coupling; see Figure \ref{fig:BF} (b). Due to the coupling $\gamma_2$ and $\gamma_4$, the near horizon BF bound can be non-violating for the large value of $Q/\kappa^2$. It indicates the possibility of the zero-temperature transition between the normal and superconducting phases. Moreover, if the scalar mass is in the range of $-3/4 < m^2 <0$, then the near horizon BF bound is violated in a finite range of $Q/\kappa^2$ (Dashed line in the figure, for example). Therefore, superconducting domes can be constructed in the full range of charge carrier density and temperature space. However, this is only the analysis from BF bound. To realize the superconducting dome in the parameter space, we have to numerically solve full equations of motion with proper boundary conditions.

\section{Background solution and phase diagram}

In the previous section, we discussed the possibility of constructing a superconducting dome by using the BF bound analysis. But it is only applicable in zero temperature limit. To see the whole phase structure in the parameter space, we have to solve equations of motion directly. There is one more comment on the BF bound analysis. In our numerical experience, the phase transition density at zero temperature is not exactly the same as BF bound analysis. The actual transition points are usually located inside the boundary of BF bound analysis in Figure \ref{fig:BF}.

In this section, we numerically solve the equations of motion(\ref{eom1}) with proper boundary conditions. We want to construct background geometry with a black hole horizon. Then, we can impose the metric function and the gauge field vanishes at the black hole horizon as
\begin{align}\label{horizon}
U(r) \Big|_{r\rightarrow r_h} &\sim (r-r_h) U'(r_h)  \cr
a_t (r) \Big|_{r\rightarrow r_h} &\sim (r - r_h) a_t'(r_h),
\end{align}
and the scalar field $\phi(r)$ and wapping factor $W(r)$  go to finite value. For the numerical calculation, we fix $16\pi G=L=1$ from now.

Together with (\ref{horizon}) and the equations of motion (\ref{eom1})-(\ref{eom4}), we get near horizon behavior of each field by the regularity condition as
\begin{align}\label{rhcond}
&a_t (r_h) =0,~~~~~~~a_t'(r_h) = a_h' \cr
&\phi (r_h) = \phi_h,~~~~~~\phi'(r_h) = \frac{2\,  \phi_h (2 m^2 - \gamma_2 \, a_h'^2 +2 \gamma_4  \,a_h'^4)}{2(6 - m^2 \phi_h^2 -\kappa^2) - a_h'^2 (1+\gamma_2\, \phi_h^2) + 6 \gamma_4 \,\phi_h^2 \,a_h'^4} \cr
&U(r_h) = 0,~~~~~~ U'(r_h) = \frac{1}{4}\left( 12 - a_h'^2 -2 m^2 \phi_h^2  - 2\kappa^2 - \gamma_2 \, \phi_h^2 \, a_h'^2 + 6 \gamma_4 \, \phi_h^2 \, a_h'^4 \right).
\end{align}
Notice that the solution of the coupled nonlinear equations (\ref{eom1}) - (\ref{eom4}) can be classified by the horizon value of the scalar field $\phi_h$ and the derivative of the $U(1)$ gauge field $a_h'$ only. 

\subsection{Superconducting Dome}
In this section, we discuss the non-linear electrodynamic interaction effect on the transition between the normal phase and the hairy black hole phase, which is interpreted as a superconducting phase in the boundary system.

The most simple solution of the model is the vanishing scalar field solution. In the absence of the scalar field, all the non-linear electrodynamic interaction terms vanish and the action becomes that of the usual Einstein-Maxwell-Axion system.
The solution of this system is nothing but the RN black brane solution with momentum relaxation given by
\begin{align}
U(r)&=r^2-\frac{\kappa^2 }{2}-\frac{M }{r}+\frac{{\cal Q}^2 }{4 r^2}\,, \cr
A_t(r)&=\mu  -\frac{{\cal Q} }{r}, \cr
\phi(r)&=W(r)=0\,.
\end{align}
For the RN black brane, the mass parameter can be written by
\begin{align}
M= r_h^3+\frac{ {\cal Q}^2}{4 r_h}-\frac{\kappa ^2 r_h}{2 }\,,
\end{align}
where $r_h$ is the location of the horizon. The hawking temperature and the entropy density are
\begin{align}
T = \frac{1}{4 \pi  } \left(3 r_h-\frac{ {\cal Q}^2}{4 r_h^3}-\frac{\kappa ^2 L^4}{2 r_h}\right)~,~s = 4 \pi r_h^2\,.
\end{align}
The corresponding phase of this system is believed to be a metallic phase. The electronic properties of the system are widely studied in many literatures.

The equations of motion(\ref{eom1}) also give a solution of hairy black brane solution with asymptotic AdS as
\begin{align}
U(r) \Big|_{r \rightarrow \infty} \sim r^2 +\cdots, ~~~~~W(r) \Big|_{r \rightarrow \infty} \sim W_0 + \cdots.
\end{align}
Due to the non-linear electrodynamic interaction, there is no conserved charge. Instead, we have to read off charge carrier density and conjugate chemical potential from the asymptotic expansion of the gauge field;
\begin{align}
A_t (r) \Big|_{r \rightarrow \infty} \sim \mu - \frac{{\cal Q}}{r} +\cdots,
\end{align}
where the physical charge carrier density ${\cal Q}$ is obtained by ${\cal Q} = e^{-W(\infty)} a_t'(\infty)$. With these asymptotic behaviors of the metric and gauge field, the scalar field can be expanded near the boundary,
\begin{align}\label{aymphi}
\phi (r) \Big|_{r \rightarrow \infty} \sim \frac{{\cal O}_{\phi}^{-}}{r^{\Delta_{-}}} + \frac{{\cal O}_{\phi}^{+}}{r^{\Delta_{+}}} +\cdots, ~~~~~\Delta_{\pm} = \frac{3}{2} \pm \sqrt{\frac{9}{4} +m^2}.
\end{align}
As a traditional holographic superconductor model, ${\cal O}_{\phi}^{\pm}$ can be interpreted as a source and vacuum expectation value of the corresponding operator. In this work, we consider ${\cal O}_{\phi}^{-}$ as a source which is vanishing for the spontaneous condensation.

\begin{figure}[ht!]
\begin{center}
\subfigure[$\gamma_2 =6,~\gamma_4 =1.5$]
   { \includegraphics[width=7cm]{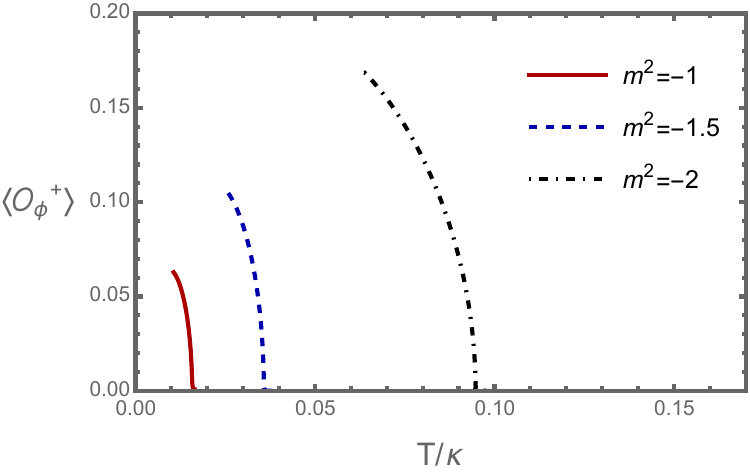}  }
   \hskip1cm
 \subfigure[$\gamma_2=6,~m^2= -1$]
   { \includegraphics[width=7cm]{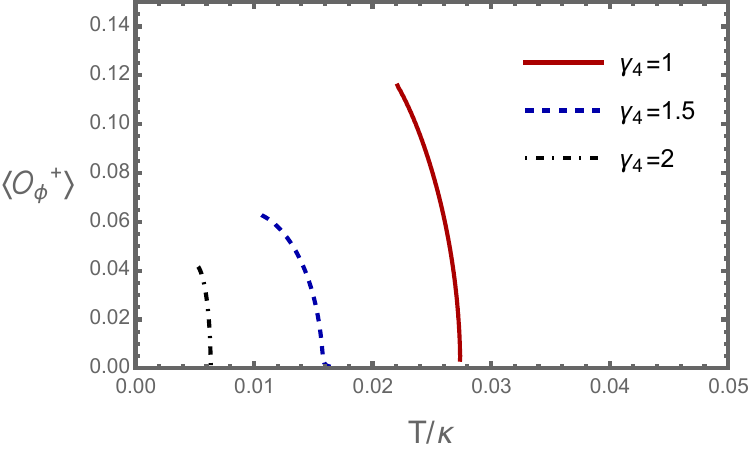}  }

\caption{Temperature dependence of the value of the scalar field for (a) fixed $\gamma_2$ and $\gamma_4$, (b) fixed $\gamma2$ and $m^2$. We fix the scalar charge to $e =1$ in both cases.}
\label{fig:TO}
\end{center}
\end{figure}

Numerical calculations show that there is a condensation of the scalar field without a source in the presence of the non-linear electrodynamic coupling, see Figure \ref{fig:TO}. In the figure, each line denotes a value of the scalar condensation and the line continues along the $T/\kappa$ axis which indicates a normal RN AdS black hole. Figure \ref{fig:TO} (a) shows temperature dependences of the scalar condensation for different scalar mass $m^2$. As the scalar mass decreases, the transition temperature increases. It can be understood that the difference between the scalar mass and near horizon BF bound increases and hence, the scalar field can easily condensed. The temperature dependences of the scalar condensation for different $\gamma_4$ interactions are drawn in Figure \ref{fig:TO} (b). As shown in the figure, the $\gamma_4$ interaction term tends to decrease transition temperature. It is also consistent with BF bound analysis in (\ref{BFcond}) because the $\gamma_4$ interaction term increases the effective mass. The effect of the $\gamma_2$ interaction can be easily expected by (\ref{BFcond}). This term is always lowering the effective mass of the scalar field, therefore the transition temperature will increase. We do not show the case here, but we checked it numerically.

\begin{figure}[ht!]
\begin{center}
\subfigure[$m^2 =-2,~\gamma_2 =\gamma_4 =0$]
   { \includegraphics[width=6.5cm]{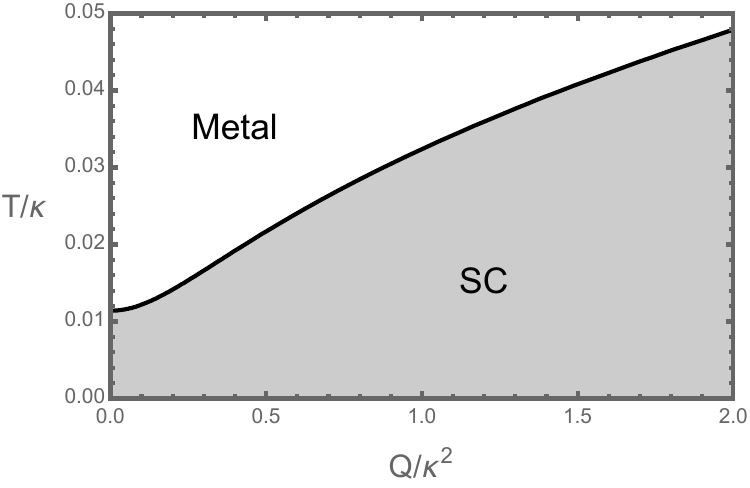}  }
   \hskip1cm
 \subfigure[$m^2=-0.5,~\gamma_2=0.5,~\gamma_4=0$]
   { \includegraphics[width=6.5cm]{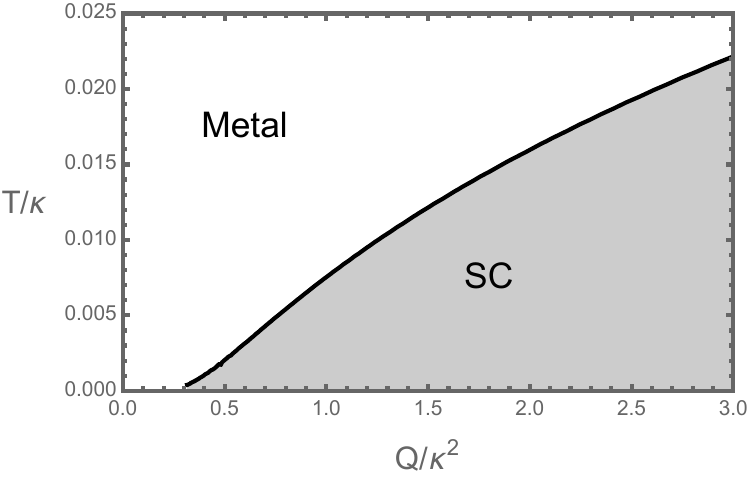}  }
  \subfigure[$m^2=-1.5,~\gamma_2=6,~\gamma_4=1.5$]
   { \includegraphics[width=6.5cm]{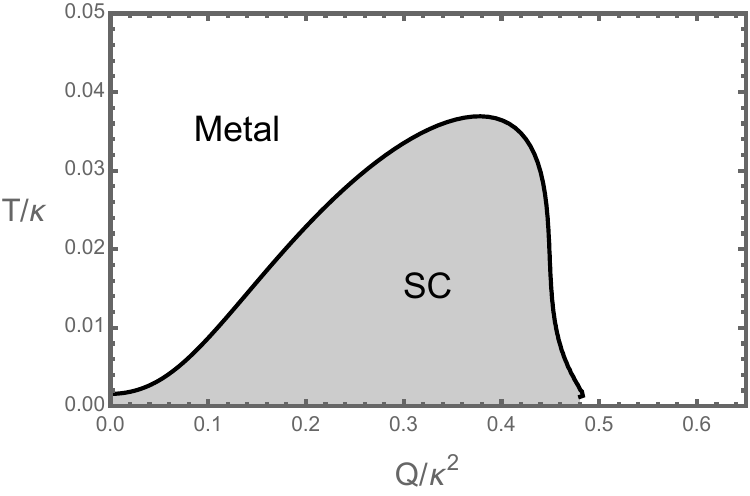}  }
   \hskip1cm
  \subfigure[$m^2=-1,~\gamma_2=6,~\gamma_4=1.5$]
   { \includegraphics[width=6.5cm]{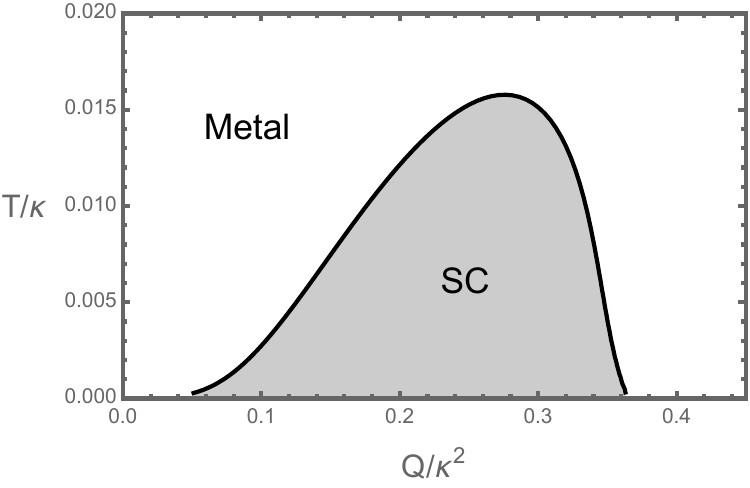}  }

\caption{Non-linear electrodynamic interactions and scalar mass dependence of phase diagram. We fix the scalar charge to $e =1$ in all cases.}
\label{fig:PTs}
\end{center}
\end{figure}

Depending on the non-linear electrodynamic interaction $\gamma_2$ and $\gamma_4$, the superconducting phase diagram changes dramatically as shown in Figure \ref{fig:PTs}.
In the figure, the shaded region denotes a hairy black hole region, which corresponds to the superconducting phase and the white region is RN AdS black hole hence, we can identify this region as the metallic phase.   Figure \ref{fig:PTs} (a) corresponds to the phase diagram of an ordinary holographic superconductor. One can see that the low temperature region is always covered by the superconducting phase.

We choose four different sets of parameters in Figure \ref{fig:PTs} such that they clearly show four classes of the phase structure. In the figure, we scale temperature $T$ and charge carrier density ${\cal Q}$ with impurity density $\kappa$ such that both axes are dimensionless. The classification of the phase diagram is as follows: \\\\
Figure \ref{fig:PTs} (a): In this case, the superconducting phase covers the whole low temperature region. Transition temperature increases as charge carrier density increases. There is a phase transition even in zero charge density. \\
Figure \ref{fig:PTs} (b): Superconducting phase occupies low temperature and large charge carrier density region. But there is no superconducting phase in small charge carrier density. It implies there is a phase transition at zero temperature. We can say there is a quantum phase transition from the metallic phase to the superconducting phase as the charge density increases. \\
Figure \ref{fig:PTs} (c): In this case, the large charge carrier density region is fully occupied by the metallic phase. But low temperature and small charge density region(including zero charge density) is covered by the superconducting phase. We can say there is a quantum phase transition from the superconducting phase to the metallic phase. \\
Figure \ref{fig:PTs} (d): This phase diagram shows superconducting phase exists in a finite range of charge carrier density and temperature, the so-called `superconducting dome'. In this case, there are two quantum phase transition points.

\begin{figure}[ht!]
\begin{center}
   { \includegraphics[width=5cm]{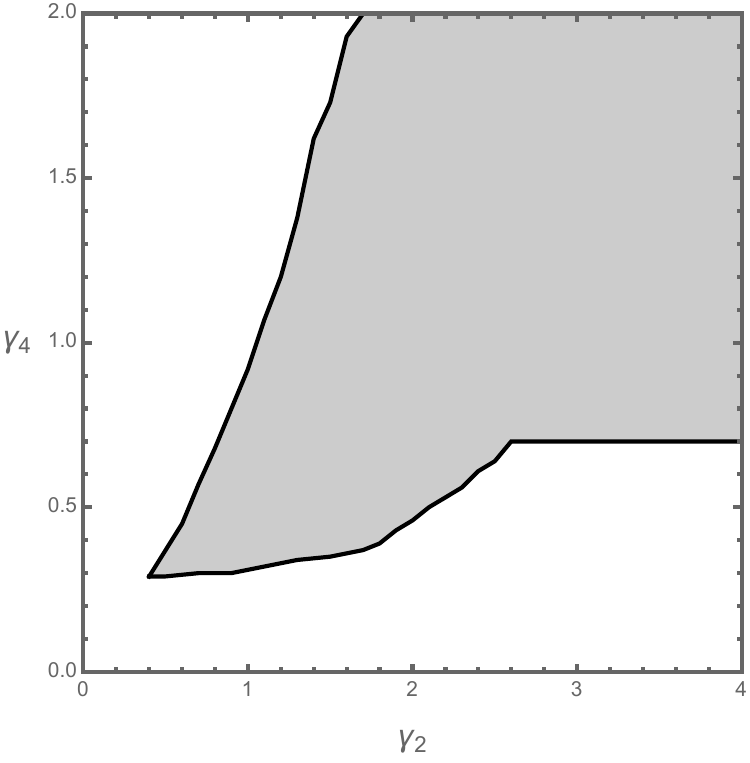}  }
  
\caption{The region of parameters $\gamma_2$ and $\gamma_4$ where the superconducting dome exist. We fix the scalar charge to $e =1$ and the scalar mass $m^2=-1$.}
\label{fig:domepara}
\end{center}
\end{figure}

Figure \ref{fig:domepara} shows the possible range of non-linear electrodynamic interactions $\gamma_2$ and $\gamma_4$ where the superconducting dome forms in the phase diagram. We fix the mass of the scalar field $m^2$ to be $-1$ and the charge of the scalar field $e=1$. The shaded region in the figure denotes the parameter space where the superconducting dome exists. In the figure, it is clear that we need not only $\gamma_2$ interaction but also $\gamma_4$ term to construct a superconducting dome. This is already speculated in BF bound analysis (\ref{BFcond}).  As we change the mass, the parameter region for the superconducting dome would be changed. But it only can be checked by numerical calculations and we show a single case in this paper. 

In this work, we are mainly focused on the construction of the superconducting dome. Therefore, we will concentrate on the Figure \ref{fig:PTs} (d) case. We postpone the details of other cases in future work.

\subsection{New scaling geometry}

In this section, we study background solutions in the superconducting phase. To do this, we choose parameters in the model used in Figure \ref{fig:PTs} (d). As temperature decreases, one can expect the value of condensation will increase, satisfying source free conditions as shown in Figure \ref{fig:TO}. However, when we lower the temperature for a given ${\cal Q}/\kappa^2$ line, we cannot get source source-free solution of the scalar field below a certain temperature.

\begin{figure}[ht!]
\begin{center}
\subfigure[]
   { \includegraphics[width=5.5cm]{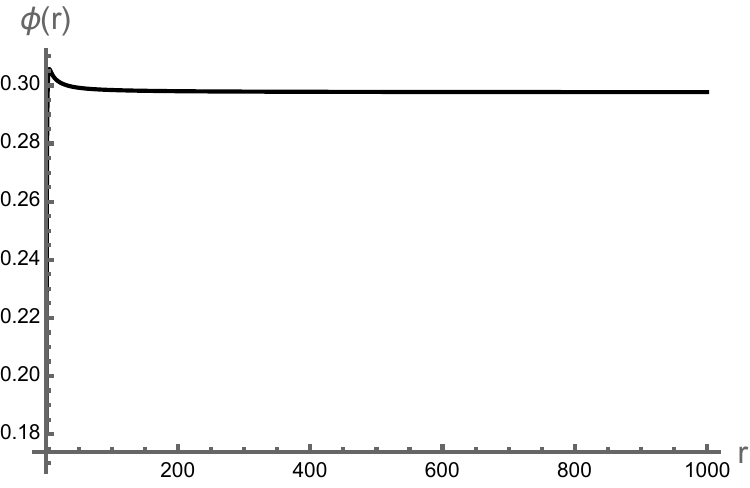}  }
\hskip1cm
 \subfigure[]
   { \includegraphics[width=5.5cm]{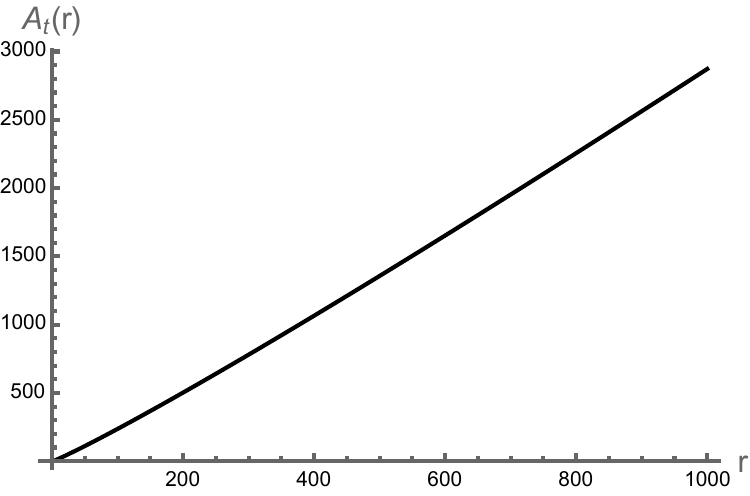}  }
 
\caption{Numerical solution of (a) scalar field $\phi(r)$ and (b) gauge field $A_t (r)$ in low temperature region. We use parameters in Figure \ref{fig:PTs} (d)}
\label{fig:Lifsol}
\end{center}
\end{figure}
The numerical solutions of each field are totally different from the solutions of hairy black hole. See Figure \ref{fig:Lifsol}. Instead of vanishing at the boundary, the scalar field goes to a finite value. The gauge field monotonically increases while it goes to the value of the chemical potential in the asymptotic AdS cases. We also check that the metric function $U(r)$ increases as $r^2$ in the asymptotic region.

The behavior of the scalar field and the $U(1)$ gauge field obviously implies that the asymptotic geometry cannot be an AdS spacetime. To see more details of the background geometry, we substitute the asymptotic behavior of the fields

\begin{align}
\phi(r) \rightarrow \phi_0, ~~~~~~~ U(r) \rightarrow  \frac{r^2}{L_0^2},
\end{align}
to the equations of motion (\ref{eom1})-(\ref{eom4}), then, we find that the solution satisfying equations of motion as 
\begin{align}
A_t(r) \sim  A_0 \,r^z, ~~~~~~W(r) \sim \frac{1}{2} \log \left(\frac{e^2 A_0^2 \phi_0^2 L_0^4\, r^{2(z-1)}}{2 (z-1)}\right),
\end{align}
with the following conditions;
\begin{align}\label{Lifcond1}
\frac{(z-1)z^2}{e^2 L_0^4} \gamma_2 -\frac{12 (z-1)^2 z^4}{e^4 L_0^8 \phi_0^2} \gamma_4  - 6 + m^2 \phi_0^2 +\frac{2(2+z)}{L_0^2} +\frac{(z-1)z^2}{e^2 L_0^4 \phi_0^2} &=0 \cr
e^2 L_0^4 \phi_0^2 (z-1) z^2 \gamma_2-4(z-1)^2 z^4 \gamma_4 -e^4 L_0^6 \phi_0^2 (2-2z+ m^2 L_0^2 \phi_0^2) &=0 \cr
e^2 L_0^4 \phi_0^2 z \gamma_2 -8(z-1)z^3 \gamma_4 + e^2 L_0^4 (z - e^2 L_0^2 \phi_0^2) &=0.
\end{align}
Combining all asymptotic solutions, the asymptotic geometry becomes Lifshitz geometry!
\begin{align}
ds^2|_{r\rightarrow \infty} 
&\sim -\frac{A_0^2 e^2 L_0^2 \phi_0^2}{2(z-1)}\, r^{2z} dt^2 + \frac{L_0^2}{r^2} dr^2 +r^2 (dx^2 + dy^2),
\end{align}
where $z$ denotes the dynamical critical exponent in the Lifshiz scaling theory.

\begin{figure}[ht!]
\begin{center}
\subfigure[]
   { \includegraphics[width=5.5cm]{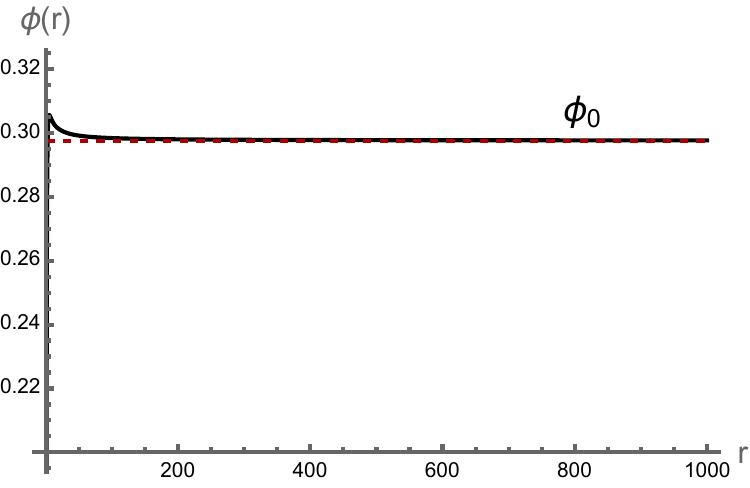}  }
\subfigure[]
   { \includegraphics[width=6cm]{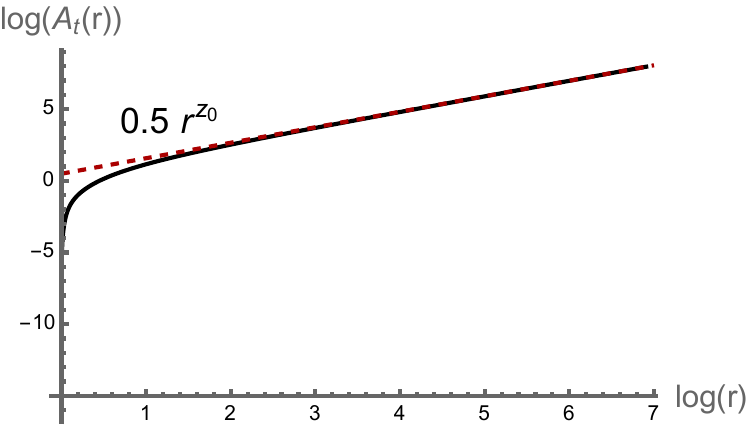}  }

\caption{Comparison of numerical results with analytic results from (\ref{Lifcond1}). (a) Comparison of the scalar field. (b) Comparison of the gauge field. We draw log-log plot for better comparison.  }
\label{fig:Asol2}
\end{center}
\end{figure}

The condition (\ref{Lifcond1}) looks very complicated. However, these equations are nothing but coupled algebraic equations, and hence we can get $(\phi_0,~z,~L_0)$ in terms of $(e,~m^2,~\gamma_2,~\gamma_4)$. For example, if we substitute parameters used in Figure \ref{fig:PTs} (d), $(e,~m^2,~\gamma_2,~\gamma_4) =(1,~-1,~6,~1.5)$ into (\ref{Lifcond1}), then we get $(\phi_0,~z_0,~L_0) \sim (0.297,~1.081,~0.939)$. Figure \ref{fig:Asol2} shows that the numerical solutions of the scalar field and the gauge field match very well with the analytic result from (\ref{Lifcond1})(red dashed line). We check other numerical solutions with different parameters and confirm that (\ref{Lifcond1}) holds for other cases.

\begin{figure}[ht!]
\begin{center}
   { \includegraphics[width=7.5cm]{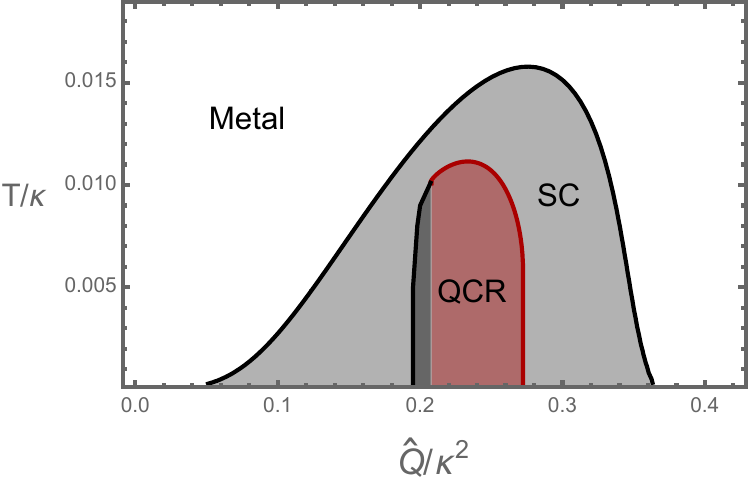}  }

\caption{Phase diagram with $m^2= -1$, $\gamma_2 = 6$, $\gamma_4 =1.5$ and $e=1$. The numerical calculations are not well solved in the dark gray region.  }
\label{fig:PTL01}
\end{center}
\end{figure}

The final phase diagram of our model is drawn in Figure \ref{fig:PTL01}. Here, we replace the charge density ${\cal Q}$ into  $\hat{{\cal Q}}\equiv e^{-W(r_h)} a_h'$ which is a slope of the gauge field at the horizon defined in (\ref{rhcond}) because the definition of charge density is not clear in the Lifshitz region.  We used parameters in Figure \ref{fig:PTs} (d) where the superconducting dome is well shaped. In the figure, the superconducting dome appears in the finite range of the charge density and low temperature region, denoted as `SC'. This superconducting phase is realized by spontaneous condensation of the charged scalar field with asymptotic AdS geometry. Outside of the superconducting dome is covered by a metallic phase, which corresponds to the usual RN-AdS black hole denoted by `Metal'.

Inside the superconducting phase, we find a phase that has new scaling symmetry. In this region, the scalar field goes to a constant in the asymptotic region and hence there is no scalar condensation. Moreover, the asymptotic geometry turns out to be Lifshitz geometry governed by the dynamical critical exponent $z$. In this case, the critical dynamical exponent is $z \sim 1.081$.  This value is close to $1$ but the corresponding asymptotic geometry is totally different. We speculate this region is closely related to the `quantum critical region(QCR)'.

We check that the transition between the metallic phase and the superconducting phase is the second order from the free energy calculations. However, the transition between the superconducting phase and the quantum critical region is crossover. One can understand this crossover from the geometrical point of view. From the Maxwell equation (\ref{eom2}), the second derivative of the gauge field at the near horizon becomes

\begin{align}\label{ddAt}
A_t''(r_H) \sim -2 \frac{(1+\gamma_2 \phi_h^2 -4 \gamma_4 \phi_h^2 \,a_h'^2)\,a_h'}{1+\gamma_2 \phi_h^2 -12 \gamma_4 \phi_h^2 \,a_h'^2}.
\end{align}

The denominator of (\ref{ddAt}) gives stability condition for the gauge field fluctuation and hence the parameters should be selected such that the sign does not change.  Within this condition, the numerator of (\ref{ddAt}) can change the sign according to the value of the scalar field and the derivative of the gauge field at the horizon. In the normal and superconducting phase, the sign of the (\ref{ddAt}) is negative and the slope of the gauge field decreases to radial direction. Therefore, the profile of the gauge field goes to a constant at the boundary which is interpreted as a chemical potential of the boundary theory. We can also obtain conjugate charge density by reading off the slope of the gauge field at the boundary. On the other hand, in the quantum critical region, the second derivative changes sign, and hence the profile of the gauge field increases along a radial direction. And the resulting geometry has Lifshitz scaling symmetry. The changing sign of (\ref{ddAt}) happens continuously as we change the horizon condition, therefore, the transition would be a crossover.

\section{Discussion}

In this work, we investigate a holographic superconductor model with non-linear electrodynamic interactions. Under this interaction, we find a geometrical transition between an RN-AdS black brane solution and a hairy black brane solution. From the boundary theory point of view, this transition can be understood as a metallic to superconducting phase transition. 

The non-linear interaction between the scalar field and the $U(1)$ gauge field can make a quantum phase transition that appears at zero temperature. It can be explained by BF bound analysis. Moreover, picking on the choice of the non-linear interaction $\gamma_2$ and $\gamma_4$, we get various shapes of the phase diagram for the holographic superconductor. In a certain range of the $\gamma_2$ and $\gamma_4$ parameters, we can construct a superconducting dome in the phase diagram and numerically check that the superconducting dome appears in a wide range of $\gamma_2$ and $\gamma_4$ interactions.

As we decrease the temperature in the superconducting phase, we find a region where source free condition for the scalar field cannot be satisfied. Instead of vanishing at the boundary, the scalar field goes to a constant. From the asymptotic analysis of the equations of motion, we find the background geometry has Lifshitz scaling governed by the dynamical critical exponent $z$. This critical exponent is determined by other parameters and the asymptotic value of the scalar field. From the results of numerical calculations, this critical exponent is surprisingly the same in the whole Lifshitz region. This implies that the superconducting phase diagram and the dynamical critical exponent are fixed by given parameters like $\gamma_2$, $\gamma_4$, $m^2$, etc. 

This work shows the possibility of a superconducting dome and the emergence of new scaling geometry in the superconducting phase. We are investigating the details of the Lifshitz scaling region. In this region, a constant scalar field makes a gauge field massive, and hence, Lifshitz scaling can be realized. In the traditional Lifshitz geometry, the charge density of the theory depends on the critical dynamical exponent $z$. However, the coefficient of the gauge field, which is interpreted as charge density, seems to be a free parameter in our model. We are analyzing the analytic solutions of our model and studying the thermodynamic properties of it. The analysis of the new Lifshitz geometry will be reported soon.

The other future direction of this model is a calculation of the AC conductivity. As shown in the phase diagram, there are two specific phases: superconducting and quantum critical region. In both regions, there exists a gap in the AC conductivity. One is a so-called superconducting gap, and the other one is the Lifshitz gap. By looking at the AC conductivity, we can understand the properties and process of the gap in each phase.

\section*{Appendix}
\appendix
\section{On-shell Action}\label{OnS}

Let us find the expression for the on-shell action. Employing the UV cut-off $r=\Lambda =\infty$, one can evaluate the bulk action, the Gibbons-Hawking term and the counter terms as follows:
\begin{align}
S_{bulk} =&\frac{  \mathcal{V}}{16\pi G}\int dt \left( -\kappa^2 \int_{r_h}^\Lambda  dr e^{W(r)-W(\Lambda )}   - \frac{2}{L^2}\Lambda  U(\Lambda )  \right)\\
S_{GH} =&\frac{  \mathcal{V}}{8\pi G}\int dt \sqrt{-\gamma} K|_{r=\Lambda }=\frac{\mathcal{V}}{16\pi G L^2}\int dt \Lambda   \left(\Lambda   U'(\Lambda  )+2 U(\Lambda  ) \left(\Lambda   W'(\Lambda  )+2\right)\right)\\
S_{ct}=&\frac{\mathcal{V}}{16\pi G}\int dt \sqrt{-\gamma}\left(-\frac{4}{L} + \frac{1}{2}\gamma^{\mu\nu}\partial_\mu \chi^i\partial_\nu \chi^i   \right)_{r=\Lambda }\nonumber\\
=&\frac{\mathcal{V}}{16\pi G L^3}\int dt \sqrt{U(\Lambda  )} \left(-4 \Lambda  ^2+\kappa ^2 L^4   \right)\,
\end{align}
where the spatial volume $\mathcal{V}$ denotes $\int dx dy$ and $\gamma_{\mu\nu}$ is the induced metric defined by the following ADM decomposition of the bulk metric:
\begin{align}\label{ADM}
ds^2 = N^2 dr^2 + \gamma_{\mu\nu}\left(dx^\mu + V^\mu dr \right)\left(dx^\nu + V^\nu dr \right)\,.
\end{align}
Here we used the Einstein equation to obtain $S_{bulk}$. When the source of the scalar field vanishes, the total action becomes
\begin{align}
S_{os}=&\lim_{r\to\infty}\left(S_{bulk} + S_{GH} + S_{ct}\right) =\frac{\mathcal{V}}{16\pi G}\int dt \left( \frac{M}{L^2} + \kappa^2 \Lambda  -\kappa^2\int_{r_h}^\Lambda  e^{W(r)-W(\Lambda )} \right)\, .
\end{align}
where $M$ is determined by the asymptotic behavior of $U(r)$  as follows:
\begin{align}
U(r) = \frac{r^2}{L^2} +c_1 -\frac{ML^4}{r} + \cdots \, .
\end{align}
The last term in $S_{os}$ can be replaced by a constraint equation, which is derived from the field equations and is given by
\begin{align}
L^2 \kappa^2 e^W = \partial_r^2\left(r^2 U e^W\right) + \partial_r \left( r^2 U \partial_r\left(e^W\right) \right) -\partial_r\left( 4 r U e^W\right) + L^2 \partial_r\left(a_t Q\right)  \, ,
\end{align}
where function $Q$ is defined by
\begin{align}
Q \equiv  - \frac{r^2}{L^2} a_t' e^{-W} \left(1 + \gamma_2 |\varphi|^2 - 4\gamma_4 L^2 e^{-2W}  |\varphi|^2 a_t'^2 \right) \, .
\end{align}
By integrating both sides from black hole horizon $r_h$ to $\Lambda$ and considering the asymptotic behavior of $U(r)$ and $a_t(r)$, the Euclidean on-shell action takes the form
\begin{align}
\frac{\Omega_{os}}{\mathcal{V }} = \frac{L^2}{16\pi G} \left( 2 M-\mu {\cal Q}  \right) - s T\,,
\end{align}
where $\cal Q$ is determined by the asymptotic behavior of $U(r)$ and $a_t(r)$ as follows,
\begin{align}
U(r) = \frac{r^2}{L^2} -\frac{\kappa^2 L^2}{2} -\frac{ML^4}{r} + \frac{{\cal Q}^2 L^6}{4r^2} +\cdots  \, , 
a_t(r) =  \mu L - \frac{{\cal Q} L^3}{r} +\cdots  \, .
\end{align}

To get the energy-momentum tensor, we first calculate the extrinsic curvature tensor of (\ref{ADM}), which is given by
\begin{align}
K_{\mu\nu} =\frac{1}{2N} \left(\partial_r \gamma_{\mu\nu} -D_{\mu} V_{\nu} -D_{\nu} V_{\mu} \right),
\end{align}
where $D_{\mu}$ is a covariant derivative of the boundary metric $\gamma_{\mu\nu}$. With these variables, we can define a conjugate momentum of the boundary metric as
\begin{align}
\Pi_{\mu\nu} \equiv \frac{\delta S_{os}}{\delta \gamma_{\mu\nu}} = \sqrt{-\gamma}\left( K_{\mu\nu} - K \gamma_{\mu\nu} - 2\gamma_{\mu\nu} + G_{\mu\nu}[\gamma] -\frac{1}{2} \partial_{\mu} \chi_{i}\partial_{\nu}\chi_{i} +\frac{1}{4} \gamma_{\mu\nu} \nabla \chi_{i} \cdot \nabla \chi_{i} \right).
\end{align}
Then, the boundary energy-momentum tensor becomes
\begin{align}
\left<T_{\mu \nu}\right> = \lim_{r\rightarrow \infty} \frac{2r}{\sqrt{-\gamma}} \Pi_{\mu\nu} = {\rm diag}(2M,~M,~M).
\end{align}

\section*{Acknowledgments}
We thank Sang-Jin Sin, Miok Park, Chanyong Park and Hyun Seok Yang for their helpful discussions. This work is supported by Basic Science Research Program through NRF grant No. NRF- 2022R1A2C1010756(Y. Seo, S. Kim), NRF-2019R1A2C1007396(K. K. Kim). We acknowledge the hospitality at APCTP, where part of this work was done. YS is grateful to the long term workshop YITP-T-23-01 held at YITP, Kyoto University, where a part of this work was done.

\end{document}